# Leveraging Socioeconomic Information and Deep Learning for Residential Load Pattern Prediction

Wen-Jun Tang[1], Xian-Long Lee[1], Hao Wang[2], and Hong-Tzer Yang[1]

[1]Department of Electrical Engineering, National Cheng Kung University, Taiwan

[2]Clean Energy Institute, University of Washington, Seattle, WA 98195 USA

Emails: [1]wjtang@ee.ncku.edu.tw, n27061061@mail.ncku.edu.tw , htyang@mail.ncku.edu.tw, [2]hwang16@uw.edu

***Abstract*—Advanced metering infrastructure systems record a high volume of residential load data, opening up an opportunity for utilities to understand consumer energy consumption behaviors. Existing studies have focused on load profiling and prediction, but neglected the role of socioeconomic characteristics of consumers in their energy consumption behaviors. In this paper, we develop a prediction model using deep neural networks to predict load patterns of consumers based on their socioeconomic information. We analyze load patterns using the K-means clustering method and use an entropy-based feature selection method to select the key socioeconomic characteristics that affect consumers' load patterns. Our prediction method with feature selection achieves a higher prediction accuracy compared with the benchmark schemes, e.g. 80% reduction in the prediction error.**

## I. INTRODUCTION

Increasing deployment of Advanced metering infrastructure (AMI), e.g. smart meters, produces a large volume of residential load data, a fact which opens up new opportunities for utility companies to understand consumers' energy consumption behaviors and thus potentially improve the power system operation. In particular, utility companies are interested in demand response (DR) that elicits changes in electricity consumption of consumers through incentive mechanisms [1]. However, utility companies face a fundamental problem that the consumers' energy consumption behaviors are highly uncertain, hindering the successful implementation of DR programs. In this paper, based on the residential load data, we study the load patterns of residential consumers and how to predict load patterns using demographic and socioeconomic information.

### A. Related Work

Great efforts have been made to study the load profiling and load prediction in [2]-[6]. For example, studies in [2] developed random effects mixture models to identify clusters of load series based on a load dataset from a Canadian utility company. Kwac et al. in [3] used an adaptive K-means method to analyze representative load shapes out of a population of 220 thousand residential consumers and capture consumers' lifestyle based on their load shapes. Based on the data of home devices usage collected by a wireless power meter sensor network, Barbato et al. in [4] presented a system to forecast household devices usage and in particular an algorithm to predict what devices and when the consumers will use the next day. Studies in [5] and [6] improved the accuracy for the short-term load forecasting by considering the levels of load aggregation and weather data, respectively. Studies in [7] and [8] adopted deep neural networks for short-term load forecasting and achieved higher forecasting accuracy compared with Linear Regression, Support Vector Regression, and Weighted Moving Average.

Besides the development of algorithms for improving the performance of load profiling and load prediction, recent studies in [9]-[11] focused on how to utilize load profiling and prediction to enhance the operations of utility companies. A load forecasting strategy combining wavelet transform and artificial neural networks was presented in [9] to predict the response of residential loads to different price signals in DR programs. To facilitate data-driven grid management, a lifestyle segmentation method was developed in [10] to study consumers' load shapes and the peak hours. Kwac and Rajagopal in [11] investigated the customer selection for DR programs using consumers' energy consumption data and formulated a stochastic knapsack problem for the DR operator to select the optimal set of users.

As demonstrated in the existing studies [2]-[11], smart meter data enhance the understanding of residential load patterns and thus help utility companies in the planning and operation for modernizing power systems. However, most of the existing studies, e.g. [2]-[11], heavily relied on the data analysis of each individual's historical load profile to perform various operations. It is often costly for utility companies to deploy a massive AMI, and some households may not have been equipped with smart meters [12]. Han et al. in [12] took the first step to consider the impact of socio-economic factors of users in the power load forecasting but the discussions were limited to the peak loads and total energy consumption of users. In this paper, we aim to study richer characterizations of users' loads, e.g. load patterns, and understand the role of their socioeconomic information in the load patterns.

### B. Main Contributions

Energy consumption is a direct reflection of consumer behaviors that depend on the various factors especially demographic and socioeconomic characteristics. Prior works, e.g. in [13], identified major socioeconomic features and integrated these features to forecast the peak load and total energy consumption of consumers using random forest method. To perform better operational strategies, e.g. DR, utility companies require a more thorough characterization of load profiles, but how socioeconomic features affect consumers' load profiles is not yet well-understood.

Our work analyzes consumers' load patterns and how the load patterns are correlated with socioeconomic features, e.g. number of residents, ages, annual income, and education levels. Specifically, we use the K-means method to cluster load profiles and identify key load patterns. We further study the correlation of load patterns with a set of socioeconomic

This work was supported in part by National Applied Research Laboratories (108-3116-F-492 -002 -MY2), Taiwan.

features after correlation selection and develop a prediction model that forecasts consumers' load patterns given their socioeconomic information. The trained prediction model has the potential to serve as an effective tool for utility companies to select consumers for various operations without obtaining consumers' historical load data. We summarize our contributions as follows:

- *Feature Selection:* We develop an entropy-based feature selection algorithm to identify the most correlated socioeconomic features with load patterns, and thus reduce information redundancy we use for load pattern prediction.
- *Load Pattern Prediction:* We develop a prediction algorithm using a deep neural network (DNN) to forecast load patterns of consumers only based on their socioeconomic information.
- *Numerical Insights:* We train and test our developed algorithms based on residential data and the numerical results show that our algorithm effectively leverages the socioeconomic information and achieves up to 80% reduction in the prediction compared with benchmarks.

## II. Data Pre-Processing and Load Patterns

We acquire the residential data from the Pecan Street database [14], including load data from 433 residential consumers with a resolution of 5-minutes and their socioeconomic information in 2017. The socioeconomic information includes age of residents, annual income, educational level, and the total foot square of the household. Our data processing follows the flowchart shown in Fig. 1 for the prediction of load patterns based on socioeconomic features. We will present the data pre-processing, load pattern clustering, feature selection, and prediction model in Section II.A, Section II.B, Section III.A, and Section III.B, respectively.

### A. Data Description And Pre-processing

Since the daily energy consumption is fairly different from workdays and weekends [15]-[16], we divide the dataset into two groups for the workday and weekend per the date stamps. The incomplete daily load profiles are treated invalid and thus not included. Finally, 160 out of 433 residential consumers are selected. Since the (day-ahead) energy market is usually cleared on an hourly basis, and we acquire hourly load data of

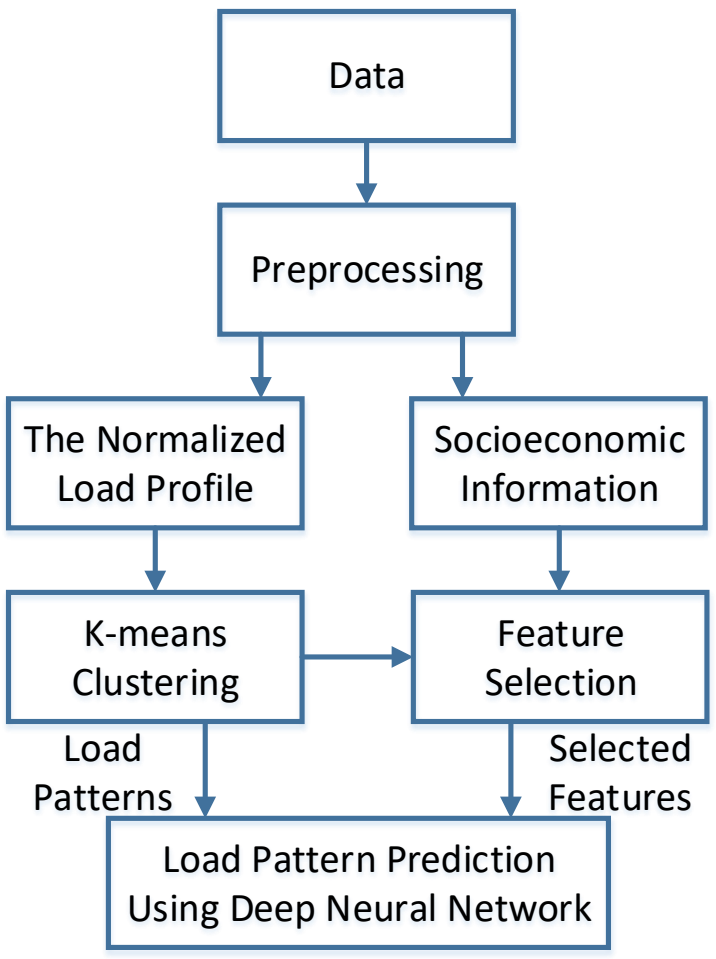

Fig. 1. Flowchart of data processing and load pattern prediction.

consumers to be consistent with the system operation. The hourly load data in workdays and weekends of user $n$ on day $d$ are denoted by $l^n_{w,d,t}$ and $l^n_{e,d,t}$, where $w$ and $e$ represent workday and weekend, respectively. Each day is divided into 24 even 1-hour intervals, i.e., $t$ =1, …, 24. Moreover, our analysis aims to capture the temporal variations and thus we normalize each consumer's hourly load profiles on each day. Specifically, we calculate the normalized workday loads $L^n_{w,d,t}$ as

$$L^n_{w,d,t} = \frac{l^n_{w,d,t} - \min_t\{l^n_{w,d,t}\}}{\max_t\{l^n_{w,d,t}\} - \min_t\{l^n_{w,d,t}\}}, \quad (1)$$

in which we normalize the hourly load profiles for each consumer over the daily load spread, which is defined as the difference between the peak load $\max_t\{l^n_{w,d,t}\}$ and minimum load $\min_t\{l^n_{w,d,t}\}$. Similarly, we compute the normalized hourly weekend load profiles $L^n_{e,d,t}$ as follows

Fig. 2.

$$L^n_{e,d,t} = \frac{l^n_{e,d,t} - \min_t\{l^n_{e,d,t}\}}{\max_t\{l^n_{e,d,t}\} - \min_t\{l^n_{e,d,t}\}}. \quad (2)$$

We label the socioeconomic information for each consumer, which includes age of residents, annual income, educational level, and the total foot square of each consumer. The related socioeconomic information is extracted into a matrix of metadata, which provides the information for the following analysis. The range of ages are sorted base on the age classification referring to the Provisional Guidelines on Standard International Age Classifications [17] as 'under 15', '15 to 24', '25 to 44', '45 to 64', and 'older than 65', respectively. The attributes are then conveyed into unique labels to further analyze the relationship between the clustered load patterns and these features.

### B. Load Patterns Clustering

Aiming at capturing the representative load patterns, we employ the K-means method [18] to categorize the daily energy consumption profile into a number of clusters, denoted as $K$. The K-means method uses the load vectors $L^n_{w,d} = \{L^n_{w,d,t}, t = 1, \ldots, 24\}$ and $L^n_{e,d} = \{L^n_{e,d,t}, t = 1, \ldots, 24\}$ (for weekdays and weekends) to calculate the minimized Euclidean distance $D$ and find the centroids over the 24 hours as

$$\text{minimize } D = \sum_{k=1}^{K} \sum_{s \in L^n_{w,d}} ||s - c_k||^2,$$

where $\{c_k,\ k = 1, \ldots, K\}$ are the calculated centroids of the load profiles and $s$ is the normalized load profile in $L^n_{w,d}$ for weekdays. When calculating the centroids for weekends, $s \in L^n_{e,d}$ is introduced instead. By choosing $K$ and iteratively computing the Euclidean distance between the load profile $s$ and re-generating centroids $c_k$, the optimal centroids $c^*_k$ are obtained with the minimized $D$ [3]. Each optimal centroid $c^*_k$ represents a typical load pattern.

The number of clusters $K$ needs to be selected to achieve a balance between the minimized distance and the clear representation of load patterns. On one hand, a smaller $D$ is often obtained when choosing a larger $K$, but centroids are often close to each other and do not exhibit meaningful difference. On the other hand, choosing a small $K$ would give representative load patterns but may results in a large $D$. Therefore, we choose $K^*$ at the 'knee point' defined in [19]. After obtaining a set of $K^*$ load patterns, we calculate the percentages $p^k_n$ for load pattern $k \in K^*$ and consumer $n \in N$.

## III. Feature Selection and Load Pattern Prediction

After clustering load patterns in Section II.B, we aim to analyze how the socioeconomic information of consumers affects their load patterns. The feature selection method is introduced to screen the redundant features. After the selection, the selected features and the percentage corresponding to each consumer's load in each pattern are treated as the input and output of a DNN model.

### A. *Socioeconomic Features Selection*

Real-world data often contain features with different correlation levels, and irrelevant or redundant features reduce the accuracy of the classification model and increase the computational burden. Therefore, feature selection serves as a key step in the classification process [20].

To find the key factors that affect load patterns, the filter models [21] work the best among existing feature selection methods. Specifically, the entropy-based class measurement [22], which has been widely employed in filter models, is utilized in our study. To minimize the redundancy, we aim to select a subset $S^k$ of features for load pattern $k$ from the original feature set $S$, i.e., $S^k \subseteq S$. The measurement is thus made to all probable subset $S'$ including various combinations of features for predicting $p_n^k$. The one with the highest measurement value is determined to be $S^k$. We define $U$ and $V$ as the indices of features in $S'$. The selected features in $S^k$ are ought to have high correlation with the predicted target $p_n^k$, but low or no correlation with each other.

Let $\mu \in U$ and $v \in V$ be random categories on the spaces of feature $U$ and $V$ with probabilities $P(\mu)$ and $P(v)$, respectively. The entropy $H(U)$ is calculated based on the marginal probability[3] $P(\mu)$ of each feature category by

$$H(U) = -\sum_{\mu \in U} P(\mu) \log(P(\mu)), \tag{3}$$

where $H(U)$ ranges from 0 to 1. The greater disorder $U$ has, a higher value $H(U)$ will result in. Similarly, the joint one, mutual information $MI(U,V)$, is obtained from the joint probability[4] $P(\mu, v)$ by

$$MI(U,V) = \sum_{\mu \in U} \sum_{v \in V} P(\mu, v) \log\left(\frac{P(\mu,v)}{P(\mu)\,P(v)}\right), \tag{4}$$

where $MI(U,V)$ is always nonnegative and is zero if and only if $U$ and $V$ are independent. A stronger dependency between $U$ and $V$ is revealed when $MI(U,V)$ is relatively large.

Moreover, the symmetric uncertainty (SU) is extended from $MI(U,V)$ by normalizing it to the entropy value of features or class labels, i.e.,

$$SU(U,V) = 2\frac{MI(U,V)}{H(U)+H(V)}, \tag{5}$$

which measures the inner correlation between features and ranges between 0-1 as well. Similarly, the target-related correlation $SU(U, p_n^k)$ is calculated. Finally, based on the calculated symmetric uncertainty, we select the subset of features $S^k$ as

$$S^k = \arg\max\left(\frac{\sum_{U \in S'} SU(U, p_n^k)}{\sqrt{\sum_{U \in S'} \sum_{V \in S'} SU(U,V)}}\right), \tag{6}$$

in which all the selected features highly correlate with the target $p_n^k$ but are less correlated with each other.

### B. *Load Patterns Prediction*

We aim to explore the relationships between the socioeconomic information and load patterns of consumers. DNN has emerged as a promising model for feature learning with many successful applications in image recognition, playing Go, and automatic translation [22]-[24]. We construct DNN models to learn the relationships between load patterns and socioeconomic information, and thus help predict load patterns especially when consumers' historical load profiles are not available. Different from existing studies in load prediction on a day-ahead or real-time basis, we take a different perspective to predict load patterns of consumers that characterize consumers' energy-consumption behaviors. We assume that the utility can collect basic socioeconomic information (e.g. ages and income) from voluntary consumers.

Specifically, we construct and train $K$ DNNs, each of which corresponds to a load pattern. Take the $k$-th DNN model as an example, we aim to map the feature in $S^k$ onto a classifier percentage $p_n^k$ as presented in Fig. 2. For each layer $l=1,..,L$, an activation function $\sigma$ is typically embedded to map the output from previous layer $o_{l-1}^k$ to the scalar state, which is also named as the output of this layer $o_l^k$. Specifically, the output $o_l^k$ for each layer $l$ is calculated based on the weight matrix $w_l^k$ between layer $l$ and layer $l$-$1$, bias $b_l^k$, and the output of previous hidden layer $o_{l-1}^k$ as

$$o_l^k = \sigma\left((w_l^k)^T\, o_{l-1}^k + b_l^k\right), \tag{7}$$

where we use a sigmoid function for the activation function. The final output $o_L^k$ of the DNN is the predicted percentage $\hat{p}_n^k$.

We randomly select a set of $N'$ households and use the corresponding data for training. We use mean squared error (MSE) to measure the discrepancy between the output $p_n^k$ and the predicted output $\hat{p}_n^k$ in

$$\mathcal{L} = \frac{1}{N'} \sum_{n'=1}^{N'} [p_{n'}^k - \hat{p}_{n'}^k]^2. \tag{8}$$

Gradient descent method is introduced to train and update the weights $w_l^k$ and bias $b_l^k$. As the training for $K$ DNN models is separated, the summation of $K$ outputted predicted percentage may not be equal to 1. Therefore, we normalize the outputs of DNNs by

$$\hat{p}'^k_n = \frac{\hat{p}_n^k}{\sum_1^K \hat{p}_n^k}. \tag{9}$$

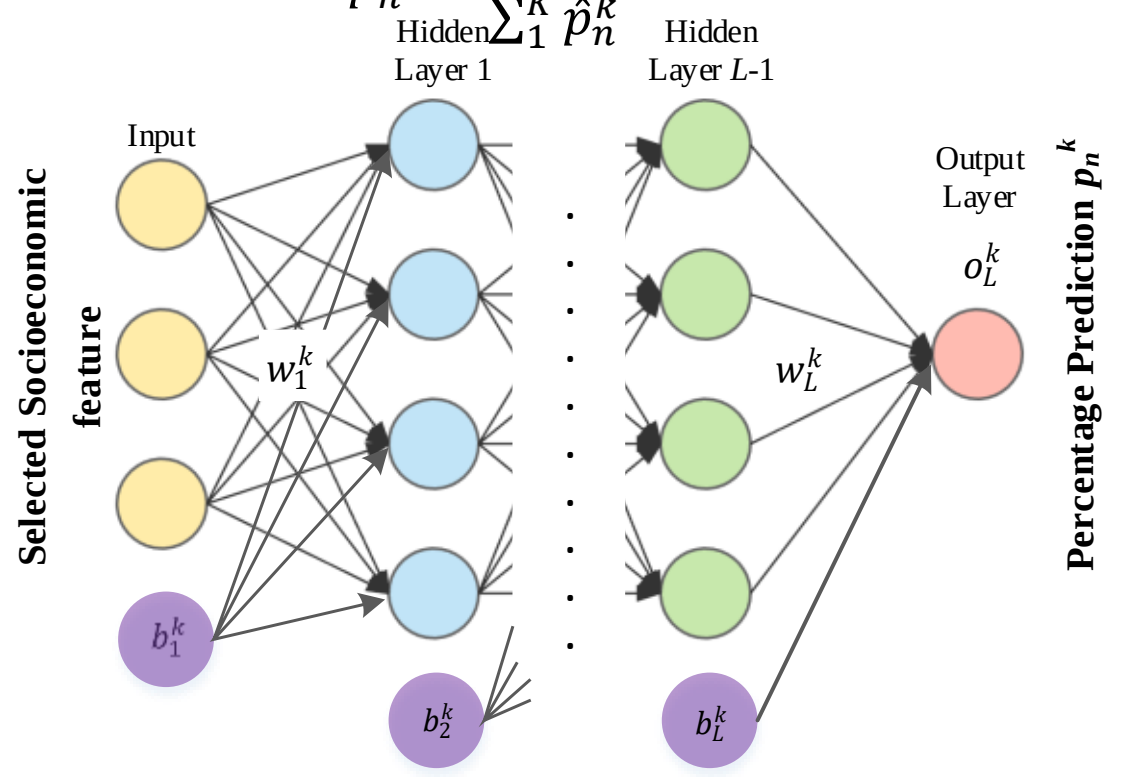

Fig. 3. The architecture of DNN.

[3] Probability of any single event occurring unconditioned on any other events.

[4] Probability of more than one event occurring simultaneously.

## IV. SIMULATION RESULTS AND DISCUSSION

To validate our methods of clustering, feature selection, and pattern prediction, we show the numerical results based on the dataset [14] described in Section II.

### *A. Load Clustering*

We randomly select 358,152 samples of the load profiles from 160 consumers, and construct 14,923 load profiles for the workday and weekend. We use the K-means method presented in section II.B to cluster load profiles and show the minimized $D$ with respect to the number of clusters $K$ in Fig. 3. We determine the value of $K^*$ using the angle-base knee point evaluation method [19] to be 7, and then generate 7 representative load patterns. We depict 7 load patterns (denoted as G1-G7) for workday and weekend, respectively in Figure 4. Each load pattern is associated with a probability that shows the percentage of consumers' daily load profiles belonging to the corresponding load pattern in the pie charts.

For the load patterns in workdays, we see that G3 and G5 are the most representative load patterns with around 40% coverage of load profiles in total. Similarly, G3 and G5 have the off-peak consumption during 13:00-18:00 and the peak after 18:00, but G3 contributes a peak around 11:00. In contrast, G1, G4, and G7 are relatively rare consumption types in workdays which all have an around 10% coverage. G1, G4, and G7 have only one peak occurring in the evening, in the morning, and early in the morning. Other two load patterns, G2 and G6, exhibit similar percentages of about 15% with a more smooth energy usage during daytime. Similar load patterns can be found in the weekend. However, there are some differences between workday and weekend load patterns. For example, comparing G1 in workdays and weekends, the load profile in the weekend is flattened and its peak is shifted 3 hours earlier than that in the workday.

### *B. Feature Selection*

Based on the feature selection method presented in Section III.A, we measure the subsets of socioeconomic features of 14 clusters (including 7 for the workday and 7 for the weekend). Each subset $S^k$ is screen from $S$ which contains all the features listed in the first column in TABLE I. The features of $S^k$ are selected by Equation (6) and checked by '$\checkmark$' in TABLE I.

We see that the feature 'Total Square Footage' does not show a strong correlation with consumers' load patterns, because load profiles have been normalized as we focus on the temporal variation of load. In contrast, age plays a key role in consumption behaviors and in particular the age greater than 65. It can be observed from G2, G3, G5, and G6 in the workday that consumers with the age greater than 65 has a high correlation with the top 4 representative load patterns. The above observation suggests that the residents whose age is greater than 65 may make a great influence on workday load patterns and are potentially the target participants in demand response programs.

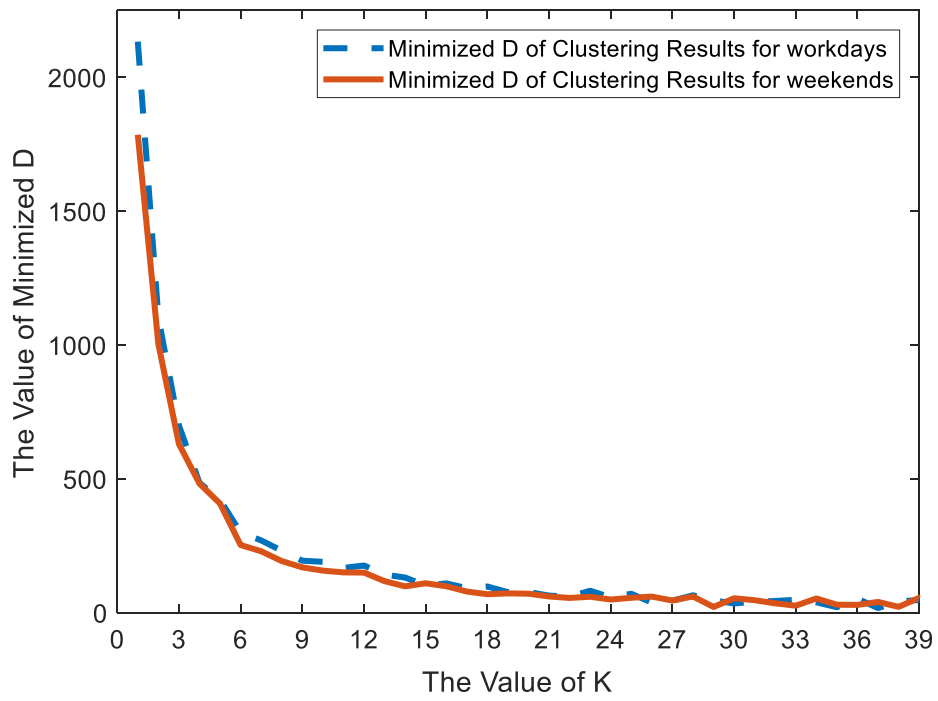

Fig. 4. The value of minimized $D$ with respect to number of clusters $K$.

### *C. Prediction Results*

The features screened in Section IV.B are taken as the inputs, and the corresponding percentages of load patterns are the outputs in our DNN models. The prediction models are separated for the workday and weekend. For both workday and weekend models, 70% of the dataset is used for training and 15% for validation. The remaining 15% of the dataset is reserved for testing.

We compare our proposed DNN model with two DNN-based models as benchmarks. In all the compared DNN models, we use the same architecture shown in Figure 2 with 10 hidden layers. Specifically, in the first benchmark, we adopt one single DNN to predict all 7 load patterns using non-selected features as inputs. The first benchmark also serves as a baseline model for comparison with our proposed prediction model that trains 7 separate DNNs for 7 load patterns, respectively. We take our model (with 7 separate DNNs for 7 load patterns) using features (without selection) as the second benchmark, denoted as "w/o Selection." Comparing with the second benchmark, we aim to validate the performance of our model denoted as "w/ Selection", when the feature selection method is used.

We use MSEs of testing results to measure the prediction performance of all compared models and the results are shown in Table II. We see that the models containing 7 DNNs regardless of feature selection outperform the baseline method with a single DNN. For example, our model without feature selection can achieve 14% and 60% reduction in MSEs on average for workday and weekend load patterns, compared with the baseline. Our model with feature selection can further reduce MSEs by 65% and 81% on average for workdays and weekends, compared with the baseline. Moreover, 60% and 54% of reductions in MSEs are achieved when using feature selection compared with the case without feature selection for the weekday and weekend, respectively.

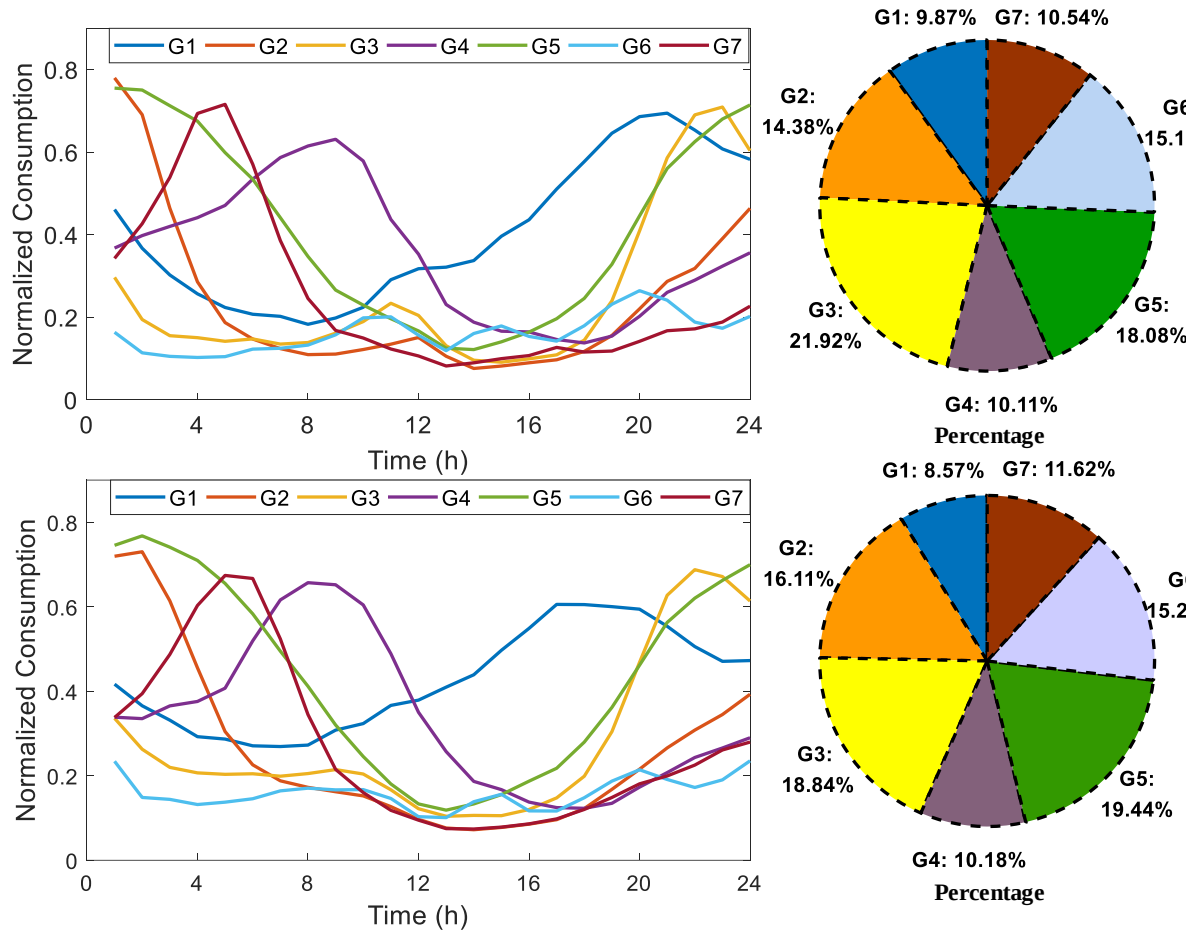

Fig. 5. The clutering results with k=7 and the corresponding ratio of load clustering.

TABLE I. THE SELECTED FEATURES FOR EACH LOAD PATTERN

| Feature | | Workday | | | | | | | Weekend | | | | | | |
|---|---|---|---|---|---|---|---|---|---|---|---|---|---|---|---|
| | | G1 | G2 | G3 | G4 | G5 | G6 | G7 | G1 | G2 | G3 | G4 | G5 | G6 | G7 |
| Age Range | under 12 | √ | √ | | | | | | √ | | √ | | | | |
| | 13-24 | √ | | | √ | | √ | | √ | √ | √ | √ | | | √ |
| | 25-49 | | | | | √ | | | | | | | √ | | √ |
| | 50-64 | √ | √ | | √ | | | √ | √ | √ | | √ | | | √ |
| | 65 and older | | √ | √ | | √ | √ | | √ | | √ | √ | | √ | √ |
| Education Level | | | | | √ | | | | | | √ | | √ | | √ |
| Annual Income | | | √ | | | | | √ | √ | | | | | √ | |
| Total Square Footage | | | | √ | | | | | | | | | | | |

TABLE II. PERFORMANCE COMPARISONS OF PREDICTION METHODS WITH AND WITHOUT FEATURE SELECTION

| Independent DNN | WORKDAY | | WEEKEND | |
|---|---|---|---|---|
| | w/o Selection | w/ Selection | w/o Selection | w/ Selection |
| G1 | 0.00911 | 0.00501 | 0.01320 | 0.00407 |
| G2 | 0.01690 | 0.00714 | 0.01130 | 0.00491 |
| G3 | 0.02420 | 0.00867 | 0.01310 | 0.00692 |
| G4 | 0.01420 | 0.00627 | 0.00342 | 0.00279 |
| G5 | 0.01610 | 0.00588 | 0.01220 | 0.00984 |
| G6 | 0.02690 | 0.00910 | 0.02880 | 0.00864 |
| G7 | 0.00769 | 0.00404 | 0.00634 | 0.00346 |
| Average MSE | 0.01644 | 0.00659 | 0.01262 | 0.00580 |
| Baseline | 0.01910 | | 0.03100 | |

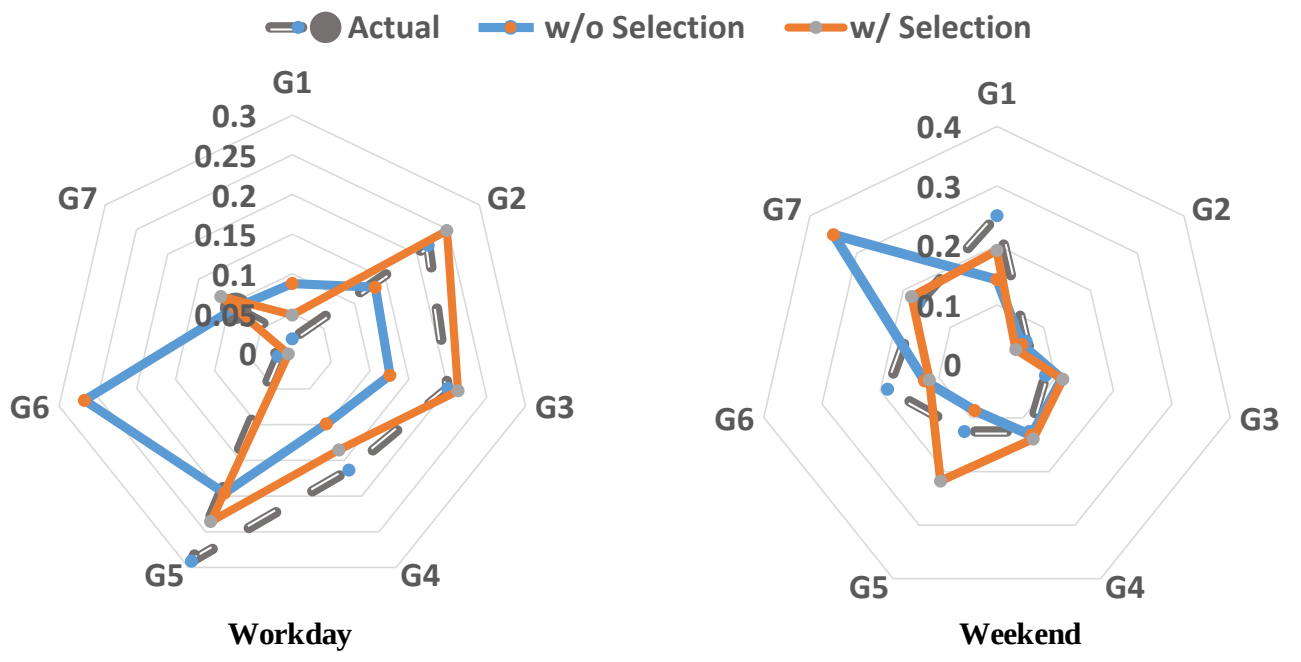

Fig. 6. The prediction result of the consumer #2814.

In Figure 5, we depict the prediction results using our DNN model with and without feature selection for one consumer (with ID #2814 in the database [13]) in the workday and at the weekend, respectively. The black dash line shows the actual value of the percentages of each load pattern. We see that without feature selection, the prediction results deviate a lot from the actual value, while the results with feature selection (shown in orange solid line) match the actual value much better. The test results demonstrated that the feature selection effectively improves the prediction accuracy.

## V. CONLCUSION

To study how energy consumption behaviors of consumers are correlated with various factors, especially socioeconomic characteristics, we developed an analytical tool that leverages deep learning techniques using a realistic load dataset. We used the K-means clustering method to identify representative load patterns among a set of consumers. We developed an entropy-based feature selection algorithm to screen a number of socioeconomic features that affect load patterns the most. Then we introduced a DNN model to predict the load patterns of consumers on workdays and in weekends based on their socioeconomic information. We discussed the numerical results of load clustering, feature selection, and the prediction results, which all demonstrated the effectiveness of our developed methods. In our future work, we will study consumer selection in demand response programs by leveraging socioeconomic information of consumers.